\documentclass[aps, twocolumn, superscriptaddress]{revtex4-1} 
\usepackage{bbm}

\usepackage{amsmath}
\usepackage{empheq}
\usepackage[parfill]{parskip}
\usepackage{graphicx}
\usepackage[active]{srcltx}
\usepackage{color}
\usepackage{array}
\newcolumntype{P}[1]{\rangle{\centering\arraybackslash}p{#1}}
\usepackage{amsfonts}
\usepackage{dsfont}
\usepackage{float}
\usepackage{subcaption}
\usepackage{lineno}
\AtBeginDocument{%
  \setlength{\linenumbersep}{.075in}%
}

\begin{document}
\setlength{\parindent}{0.5cm}

\title{Travel distance, frequency of return and the spread of disease}

\author{Cate Heine*}
\affiliation{Senseable City Lab, Massachusetts Institute of Technology, Cambridge, MA 02139} 

\author{Kevin P. O'Keeffe}

\affiliation{Senseable City Lab, Massachusetts Institute of Technology, Cambridge, MA 02139}

\author{Paolo Santi}
\affiliation{Senseable City Lab, Massachusetts Institute of Technology, Cambridge, MA 02139}
\affiliation{Istituto di Informatica e Telematica del CNR, Pisa, ITALY}

\author{Li Yan}
\affiliation{Senseable City Lab, Massachusetts Institute of Technology, Cambridge, MA 02139}

\author{Carlo Ratti}
\affiliation{Senseable City Lab, Massachusetts Institute of Technology, Cambridge, MA 02139} 

\maketitle
\section*{Abstract}
\textbf{
In 2020 and 2021, the spread of COVID-19 was globally addressed by imposing restrictions on the distance of individual travel. Recent literature has uncovered a clear pattern in human mobility that underlies the complexity of urban mobility: $r \cdot f$, the product of distance traveled $r$ and frequency of return $f$ per user to a given location, is invariant across space. This paper asks whether the invariant $r\cdot f$ also serves as a driver for epidemic spread, so that the risk associated with human movement can be modeled by a unifying variable $r\cdot f$. We use two large-scale datasets of individual human mobility to show that there is in fact a simple relation between $r$ and $f$ and both speed and spatial dispersion of disease spread. This discovery could assist in modeling spread of disease and inform travel policies in future epidemics---based not only on travel distance $r$ but also on frequency of return $f$.} 

\section*{Introduction} 
The global impact of the COVID-19 pandemic on both human health and socioeconomic activity has brought to light the importance of a nuanced understanding of the way that epidemics spread in cities. Urban spread of disease is inherently tied to human mobility: as we move through and between cities, we serve as vectors that allow disease to spread to new individuals and communities. The nature of the relationship between mobility and spread of disease has been extensively studied; it is widely understood that human travel is a driving force behind disease spread \cite{wilson1995travel, sattenspiel1995, dengue2015, flu2017, barmak2016modelling}. For this reason, many public policy interventions implemented worldwide to contain the spread of COVID-19 focused on limiting mobility, restricting the radius that individuals could travel from their homes and limiting travel between cities and countries.

As we continue to learn more about the ways that humans move, it is important to apply new  discoveries about human mobility to the study of disease spread, deepening our understanding of urban epidemics for use in future public health crises. Recent research uncovered a clear, universal pattern in urban mobility: the total distance that the average visitor to a given location travels to reach it is constant across a city, unrelated to the location's overall attractiveness~\cite{Schlapfer2021}. This value, equivalent to the distance of a location to an individual's home multiplied by the number of times that they visit it over the course of some time period, can be thought of as an ``exploration velocity": it is the effective distance individuals travel towards any given location per unit of time. It is a simple but critical parameter to understanding human movement.

Given the simplicity and apparent universality of this parameter of human mobility, its relationship to disease spread is a natural next question. Many travel restrictions and recommendations during the 2020 and 2021 COVID-19 pandemic focused on confining movements to a certain geographical area---for example, restricting the radius that people are able to travel to one’s neighborhood, city, state, or country. But what if speed of disease spread depends not just on radius of travel $r$, but also on the product $v := r\cdot f$? If $v$ has additional impact on disease spread, beyond average travel radius $r$, COVID-19 policies and restrictions that focused solely on restricting radius may have left significant unmet potential for further disease mitigation. This paper investigates the relationship between exploration velocity $v := r\cdot f$ and spread of disease, opening the door to future research on the ways in which that relationship can be utilized to contain epidemics. 


\section*{Results}
\noindent\textbf{Simulating reduction of exploration velocity on real data.}
In order to understand how $v = r\cdot f$ interacts with speed of disease spread, we start with large-scale datasets of individual human movements in New York City and Dakar, Senegal, which we call $\mathcal{M}_{\text{real}}$. The datasets each consist of a set of trips for $N$ individuals over different time periods $T$, where each trip indicates a given individual moving between two locations. We then use agent based susceptible-infected (SI), susceptible-infected-recovered (SIR), and susceptible-exposed-infected-recovered (SEIR) models calibrated with estimates for COVID-19 \cite{chen2020mathematical} to simulate disease spread as agents follow the trajectories in our datasets. Aside from their unique trajectories, each individual is assumed identical. In each simulation, we vary two parameters: the maximum travel distance $r$, measured relative to each agent's home location (see Methods for details), and the travel frequency $f$, the maximum number of times each location was visited. By varying these two parameters together, we are able to manually adjust exploration velocity $v$ in the system. In practice, this means discarding any trip of length greater than $r$ and all but $f$ randomly selected trips to a given location from our datasets. The details of the spatial partitionings used as well as other simulation details are given in the Methods.

\begin{figure*}[ht]
    \centering 
\begin{subfigure}{0.45 \textwidth}
  \includegraphics[width=\linewidth]{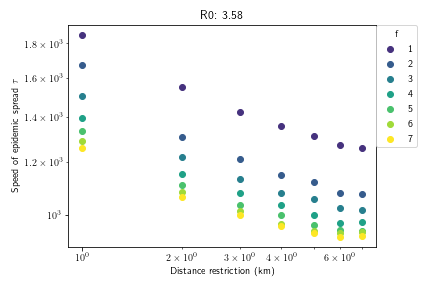}
  \caption{NYC}
  \label{rfseirnyc}
\end{subfigure}\hfil 
\begin{subfigure}{0.45 \textwidth}
  \includegraphics[width=\linewidth]{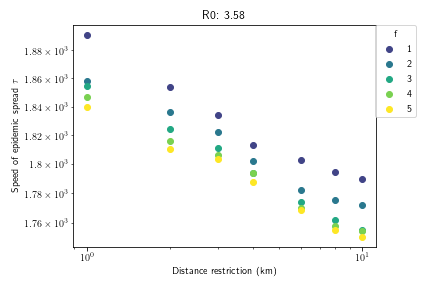}
  \caption{Dakar}
  \label{rfseirdak}
\end{subfigure}\hfil 

\medskip
\begin{subfigure}{0.45\textwidth}
  \includegraphics[width=\linewidth]{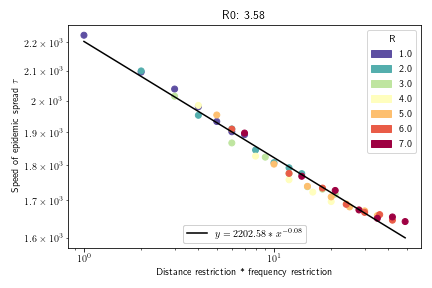}
  \caption{NYC}
  \label{rtimesfseirnyc}
\end{subfigure}\hfil 
\begin{subfigure}{0.45\textwidth}
  \includegraphics[width=\linewidth]{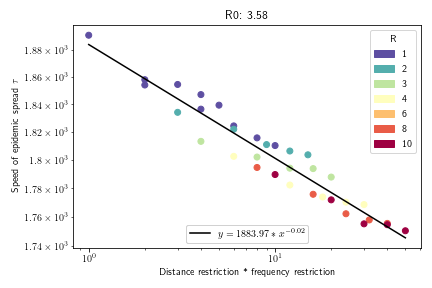}
  \caption{Dakar}
  \label{rtimesfseirdak}
\end{subfigure}\hfil 
\caption{\textbf{Epidemics sizes for SEIR model}. Top row: plots of epidemic spread speed in units of 10-minute increments against maximum allowed travel distance $r$ for different max travel frequency $f$. Bottom row: top row plotted against $r\cdot f$. Each data point is the average of 5 simulations. For other simulation details see Methods. $R^2$ values for best-fit lines are .981 (NYC) and .902 (Dakar). Best-fit line parameters are $a = -0.08, b = 2292.58$ (NYC) and  $a = -0.01, b = 1853.57$ (Dakar).}

\label{SEIR}
\end{figure*}

Figures~\ref{SEIR}(a) and (b) plot the time it takes for the disease to reach some set proportion $\tau$  of the population against exploration velocity for each dataset. The value of $\tau$ is set to 29\% for our Dakar data and 57\% for our New York City data. Thresholds are chosen to be the total epidemic size at the end of the study period under the strictest restrictions for each study area, respectively.  Simulations were run over the entire length of the datasets (14 days for the Dakar dataset; 28 days for New York City) with 10,000 agents and a 5\% initial infected population.  
The trends are intuitive and not surprising. For a given $f$, $\tau$ decreases monotonically with $r$: the further people are allowed travel, the faster the epidemic spreads. Similarly, for a given $r$, $\tau$ decreases monotonically with $f$: more frequent return trips leads to faster-spreading epidemics. What is surprising, however, is that the  $\tau(r)$ curves for each value of $f$ have similar shape. This echoes a previous finding \cite{ Schlapfer2021} and hints that $\tau, r$ and $f$ might have a simple relationship. Figures~\ref{SEIR}(c) and (d) show that they do. Under the rescaling $r \rightarrow r \cdot f$ all the data appear to merge, collapsing to a single curve that depends on the unifying factor $v := r\cdot f$, or exploration velocity. This curve can be empirically characterized as
\[
\tau(r\cdot f) = (r\cdot f)^a \cdot b
\]
 where $a$ and $b$ vary across cities and model types.
Figure~\ref{SISIR} in the Supplementary Material shows the same scaling collapse is achieved when the SI and SIR models are used, and Figure~\ref{robust} in the SM shows it persists for different values of the disease parameter $R_0$. Taken together, these findings suggest the $r \rightarrow r\cdot f$ collapse persists across multiple disease processes.

\begin{figure*}
\begin{subfigure}{0.45\textwidth}
  \includegraphics[width=\linewidth]{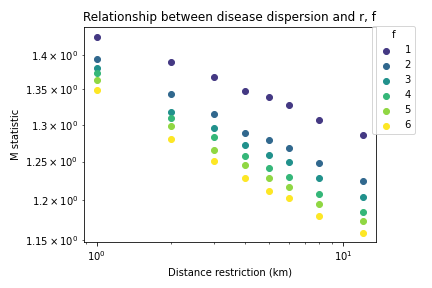}
  \caption{Relationship between frequency restriction and dispersion for radius = 700m.}
  \label{fig:spatiala}
\end{subfigure}\hfil 
\begin{subfigure}{0.45\textwidth}
  \includegraphics[width=\linewidth]{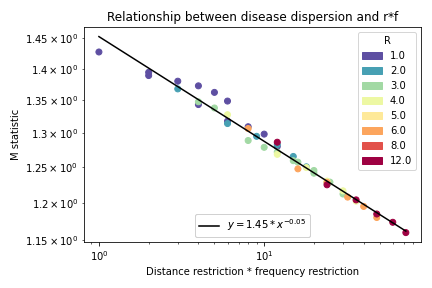}
  \caption{Scaling collapse for radius = 700m. $R^2$ value of best-fit line is .97.}
  \label{fig:spatialb}
\end{subfigure}\hfil 
\begin{subfigure}{.4\textwidth}
  \includegraphics[width=\linewidth]{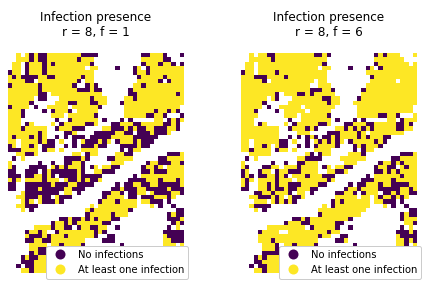}
  \caption{Holding $r$ constant but restricting $f$, we can see that loosening restrictions on $f$ increases spatial dispersion of infections---when $f$ is at 1 (left), infections reach 59.0\% of grid cells; when $f$ is relaxed to 6 (right), infections reach 73.8\% of grid cells.}
  \label{fig:grida}
\end{subfigure}
\begin{subfigure}{.5\textwidth}
  \includegraphics[width=\linewidth]{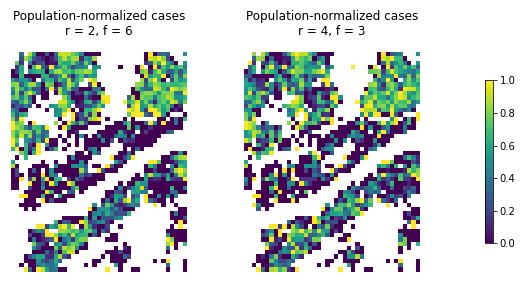}
  \caption{In the simulations above, $r$ and $f$ are individually different but $r\cdot f$ is the same. The similar spatial distribution of infections in the two simulations reflects the scaling collapse in dispersion of infection: dispersion depends on $r\cdot f$ as opposed to just $r$ or $f$. Here, color represents proportion of the grid cell population that is infected at the end of the simulation.}
  \label{fig:gridb}
\end{subfigure}\hfil 
\caption{\textbf{Scaling collapse of spatial dispersion of infections.} Like $\tau$, spatial dispersion $M$ shows a universal scaling collapse with the product $r\cdot f$, as demonstrated in panels (a) and (b). Here, $M$ is calculated using a radius of $k = 700$ meters. This is visually represented in panel (c), where we see the final spatial distribution of infections for two simulations with the same $r$ but different $f$, and in panel (d), where we see the final spatial distribution of infections for two simulations with the same $r\cdot f$ but different $r$ and $f$.}
\label{fig:spatial}
\end{figure*}

After observing the relationship between $r\cdot f$ and epidemic size, we check whether there is a relationship between $r\cdot f$ and spatial dispersion of the disease. We analyze the spatial concentration of infected persons in our New York City simulations using the $M$ function developed by Marcon and Puech, which is calculated as a function of some radius $k$ around each agent~\cite{Marcon2010}---see Methods for a detailed description of this statistic. High $M$ indicates tighter clusters of cases; low $M$ indicates a more homogeneous distribution of cases across space. We find that for a given radius $k$, as $r$ and $f$ increase, $M(k,r,f)$ decreases: infections become consistently more dispersed across the city (see Figure \ref{fig:spatiala} for this relationship when $k=700$m). As with epidemic size, under the rescaling $r \rightarrow r \cdot f$, this relationship collapses to a single curve, as shown in Figure \ref{fig:spatialb}. This indicates that restricting $r\cdot f$ is more effective than restricting $r$ alone at containing geographic spread of disease, just as with slowing epidemic spread.  This relationship is robust to the choice of $k$---as $k$ increases, $M(k,r,f)$ decreases (as would be expected---the larger radius you look at around an agent, the more representative sample of the whole population you will capture) but stays significant and maintains its relationship to $r\cdot f$, as illustrated in Figure \ref{fig:spatialcollapse} in the Supplementary Material. This relationship is demonstrated visually in Figures \ref{fig:grida} and \ref{fig:gridb}. When $r$ is fixed at 8, increasing $f$ from 1 to 6 has a clear effect on the extent to which infections spread across the city (Figure \ref{fig:grida}); while significant portions of the city remain untouched by infection when $f=1$, many more grid cells contain at least one infection when $f=6$. In Figure \ref{fig:gridb}, on the other hand, we see that when $r\dot f$ is held fixed at 12, spatial distribution of infection rates is almost identical for different $r$ and $f$ values, demonstrating the relationship between spatial distribution and $r\cdot f$.\\

\begin{figure*}
\begin{subfigure}{0.3\textwidth}
  \includegraphics[width=\linewidth]{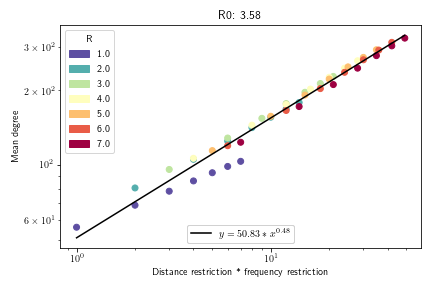}
  \caption{Relationship between $r\cdot f$ and mean degree $\langle k \rangle$.}
  \label{fig:k}
\end{subfigure}\hfil 
\begin{subfigure}{0.3\textwidth}
  \includegraphics[width=\linewidth]{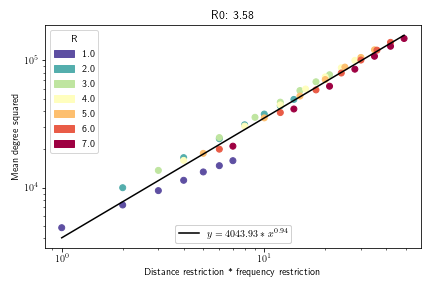}
  \caption{Relationship between $r\cdot f$ and mean degree squared $\langle k^2 \rangle$.}
  \label{fig:k2}
\end{subfigure}\hfil 
\begin{subfigure}{.3\textwidth}
  \includegraphics[width=\linewidth]{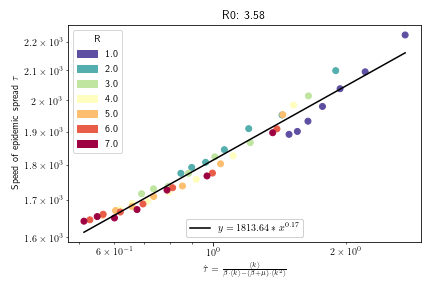}
  \caption{Relationship between simulated $\tau$ and $\tau$ as predicted by $\langle k \rangle$, $\langle k^2 \rangle$.}
  \label{fig:tautau}
\end{subfigure}
\caption{\textbf{Relationship between spreading speed and number and variance of contacts.} Both mean number of contacts $\langle k \rangle$ and mean squared number of contacts $\langle k^2 \rangle$ of the agents in our dataset show a clear logarithmic relationship with exploration velocity $r\cdot f$. The relationship between $r\cdot f$ and $\langle k \rangle$, $\langle k^2 \rangle$ can predict spreading speed $\tau$ using a simple relation that is a known result of the SIR model.}
\label{fig:k_k2_tau}
\end{figure*}

\noindent\textbf{Mechanisms.}
In order to illustrate the potential mechanism by which $r\cdot f$ determines disease spread time, we assess the relationship between $r\cdot f$ and both number and variance of contacts between agents in our simulations. Think of our simulations as a network, where agents are nodes and edges are formed whenever two agents come within some distance $\epsilon$ of one another, enabling disease transmission. It is a known result of the SIR model that speed at which a disease spreads through a network of individuals depends not only on the average degree, or number of contacts per individual $\langle k\rangle$, but also on the variance of contacts (via $\langle k^2\rangle$, or mean squared degree) according to the following relation:
$$\hat{\tau} = \frac{\langle k\rangle}{\langle k^2\rangle - (\gamma + \beta)\langle k\rangle},$$
where $\hat{\tau}$ is the characteristic time, or time it takes for the disease to reach $1/e = 36\%$ of the population, $\gamma$ is the daily recovery parameter and $\beta$ is the infection parameter. This relation isn't perfectly applicable to our context---it assumes exposure to all contacts at all timesteps, whereas our simulations incorporate movement in space and thus non-constant degree $\langle k\rangle$ over time. However, it is valuable in illustrating the effects of restricting $r$ and $f$ in our dataset. Restricting radius of travel $r$ and frequency of return $f$ affects both $\langle k\rangle$ and $\langle k^2\rangle$. The closer you stay to home and the fewer trips you take, the fewer unique individuals you have the opportunity to encounter (reducing average number of contacts $\langle k\rangle$). Further, as stricter restrictions are enforced, high-r, high-f trips are removed and the mobility patterns of high-frequency, long-distance travellers start to look more like the mobility patterns of low-frequency travellers who stay close to home (reducing variance of contacts $\langle k^2\rangle$).
Figure \ref{fig:k_k2_tau} plots these relationships in our New York City data, showing exploration velocity $r\cdot f$ on the x-axis against average number of contacts $\langle k\rangle$ and average squared number of contacts $\langle k^2\rangle$ on the y-axis in panels a and b, respectively. Again, we see a logarithmic relationship. Plugging both $\langle k\rangle$ and $\langle k^2\rangle$ into the relation above, we see that characteristic time as predicted by the degree distribution ($\hat{\tau}$) is proportional to the true characteristic time ($\tau$) that we see in our simulations -- Figure~\ref{fig:k_k2_tau}, panel c. Taken together, these findings show that the shape of the relationship between $r\cdot f$ and $\tau$ can be predicted by decreases in number and variance of contacts between agents in combination with fundamental SIR modeling results.

\begin{figure*}
\begin{subfigure}{0.45\textwidth}
  \includegraphics[width=\linewidth]{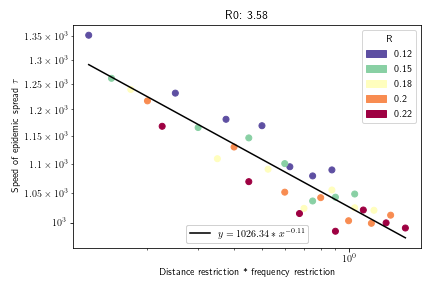}
   \caption{Relationship between distance restriction and $\tau$ in PEPR simulations with $P_{\text{travel}} = .25$. $R^2$ value of best fit line is .875.}
  \label{fig:tau_sim_100}
\end{subfigure}\hfil 
\begin{subfigure}{0.45\textwidth}
  \includegraphics[width=\linewidth]{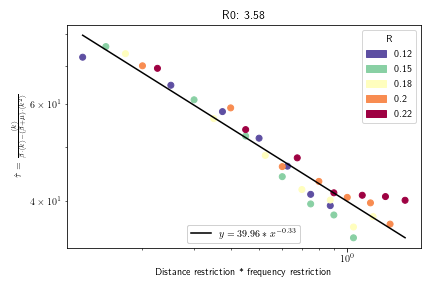}
  \caption{Relationship between distance restriction and $\hat{\tau}$ as predicted by $\langle k \rangle$, $\langle k^2 \rangle$ in PEPR simulations with $P_{\text{travel}} = .25$. $R^2$ value of best fit line is .924.}
  \label{fig:tauhat_sim_100}
\end{subfigure}
\begin{subfigure}{0.45\textwidth}
  \includegraphics[width=\linewidth]{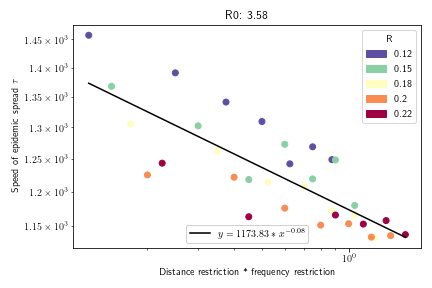}
   \caption{Relationship between distance restriction and $\tau$ in PEPR simulations with $P_{\text{travel}} = 1$. $R^2$ value of best fit line is .640.}
  \label{fig:tau_sim}
\end{subfigure}\hfil 
\begin{subfigure}{0.45\textwidth}
  \includegraphics[width=\linewidth]{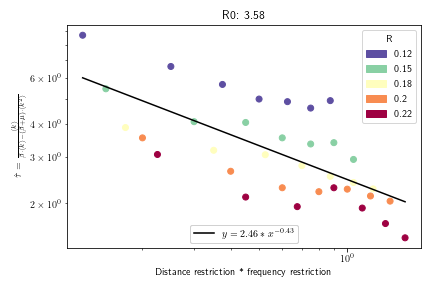}
  \caption{Relationship between distance restriction and $\hat{\tau}$ as predicted by $\langle k \rangle$, $\langle k^2 \rangle$ in PEPR simulations with $P_{\text{travel}} = 1$. $R^2$ value of best fit line is .480.}
  \label{fig:tauhat_sim}
\end{subfigure}
\caption{\textbf{PEPR simulation results.} When we run SEIR models across a set of trajectories $M_{\text{sim}}$ which have been created using the modified PEPR model with $P_{\text{travel}} = .25$, we see a similar relationship between $r\cdot f$ and $\tau$ to that in our real trajectories $M_{\text{real}}$. When we run SEIR models across trajectories which have been created using the original PEPR model (with $P_{\text{travel}} = 1$), we do not see this relationship.}
\label{fig:tau_PEPR}
\end{figure*}

\begin{figure*}
\begin{subfigure}{0.45\textwidth}
  \includegraphics[width=\linewidth]{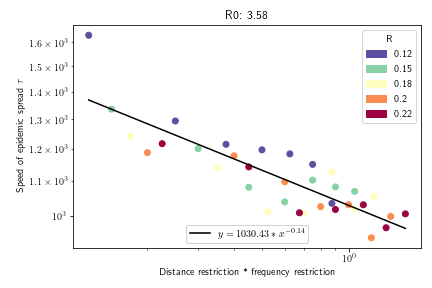}
   \caption{Relationship between distance restriction and $\tau$ in PEPR simulations.}
  \label{fig:k}
\end{subfigure}\hfil 
\begin{subfigure}{0.45\textwidth}
  \includegraphics[width=\linewidth]{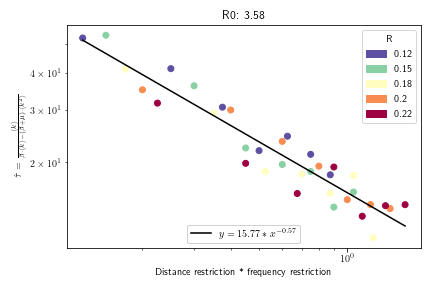}
  \caption{Relationship between distance restriction and $\hat{\tau}$ as predicted by $\langle k \rangle$, $\langle k^2 \rangle$ in PEPR simulations.}
  \label{fig:k2}
\end{subfigure}
\caption{\textbf{PEPR simulation results, where $P_{\text{travel}} = .40$.} When we run SEIR models across a set of trajectories $M_{\text{sim}}$ which have been created using the PEPR model with $P_{\text{travel}} = .40$, we see a similar relationship between $r\cdot f$ and $\tau$ to that in our real trajectories $M_{\text{real}}$.}
\label{fig:tau_PEPR_40}
\end{figure*}

These results can be replicated by a simple modification of the preferential exploration and preferential return (PEPR) mobility model presented in previous literature \cite{Schlapfer2021}. Under the PEPR model as proposed in \cite{Schlapfer2021}, at each timestep, agents either explore a new location with a fixed probability $P_{\text{new}}$ or return to a previously-visited location with complementary probability $1-P_{\text{new}}$. In order to choose a new location, agents draw a radius of travel $\Delta r$ from a heavy-tailed distribution with exponent $P(\Delta r)\sim |\Delta r|^{-1-\alpha}$, with $\alpha = 0.55$ taken from \cite{Schlapfer2021}, and they draw an angle of travel $\theta$ with preference towards $\theta$ that are heavily visited by other agents (see Methods for details). As demonstrated in \cite{Schlapfer2021}, the PEPR model replicates the universality of $r\cdot f$. However, when we run our SEIR model on mobility trajectories simulated with the PEPR model, we do not see the same relationship between $r\cdot f$ and $\tau$ that we see in our real data. This is because, in the original PEPR model, agents are always moving from place to place without necessarily returning home, whereas in our real mobility trajectories, time spent at home means less exposure to new, unique contacts (and more exposure to the same neighbors). This leads to the PEPR model being unable to replicate the changes in $\langle k \rangle$ that are associated with loosening or tightening radius restrictions, as seen in Figures \ref{fig:tau_sim_100} and \ref{fig:tauhat_sim_100}. In order to correct for this, we add a probability of travel $P_{\text{travel}}$---at each timestep, with probability $P_{\text{travel}}$, the agent travels to a new location according to the protocol above, whereas with complementary probability $1-P_{\text{travel}}$ the agent stays or returns home. Figures \ref{fig:tau_sim} and \ref{fig:tauhat_sim} show the results of rerunning the same SEIR simulations that were run on our real mobility traces $\mathcal{M}_{\text{real}}$ on a set of simulated trajectories $\mathcal{M}_{\text{sim}}$ derived from the modified PEPR model. In order to implement radius restrictions within the PEPR model we restrict the power law distribution from which $\Delta r$ is drawn and in order to implement frequency restrictions we delete all trips beyond frequency $f$ to the same location. We set $P_{\text{travel}} = .25$, the average proportion of time spent away from home in $\mathcal{M}_{\text{real}}$, and run the PEPR simulations on a unit square (see Methods for more details). We see the same logarithmic relationship between $r\cdot f$ and both $\tau$ and $\hat{\tau}$ (expected $\tau$ derived from $\langle k \rangle$ and $\langle k^2 \rangle$) that we see in our simulations over $\mathcal{M}_{\text{real}}$, although the simulated results are less precise than the results we see in $\mathcal{M}_{\text{real}}$. We also replicate the simulations with $P_{\text{travel}} = .40$ and see similar results (see Figure \ref{fig:tau_PEPR_40} in the SM). This indicates that the modified PEPR model is able to replicate the relationship between exploration velocity and speed of epidemic spread.\\

\section*{Discussion}
The significance of $r\cdot f$ was revealed in recent work \cite{Schlapfer2021} on human mobility patterns which shows the number of people who visit a given location $f$ times from a distance $r$ during a certain reference period follow a universal, inverse square law $N(r,f) \propto 1/(rf)^2$. We suspect the $\tau(rf)$ curve found in this paper derives from this inverse square law; while we have discussed the relationship between exploration velocity and network contacts, deriving the exact mechanism linking the universal inverse square law to $\tau(rf)$ is an open question for future work. 

An important implication of the $\tau(r\cdot f)$ curve is that restricting short, high frequency trips in a simulated environment does as much to slow disease spread as restricting long, low frequency trips. Agents in our simulations that take multiple trips around their neighbourhood spread disease as quickly as a those taking single, long trips to the city center. This is interesting academically, but more importantly, it has potential implications for policy: it means that restricting travel distance $r$ but ignoring travel frequency $f$ could be ineffective in containing the spreading of a disease. 

That said, an important limitation of this study is that it restricts exploration velocity in a very specific way, removing trips that extend beyond a certain $r$ or $f$. In reality, it is unclear how people would respond to policies that aim to restrict exploration velocity. True behavioral responses to different travel policies and what policies or mechanisms would be required to change individuals' behavior in a way that reduces exploration velocity are areas for future research. Further limitations of this study include the inherent limitations of simple epidemiological modeling; it is known that the SI, SIR, SEIR and other classic disease models make  assumptions which bound their accuracy \cite{cirillo2020tail,donnat2020modeling,Buchanan2020limits}. The estimates for the model parameters likely carry error \cite{ma2020epidemiological}, and we also assumed that each agent was identical and that each exposure within a set radius carried equal likelihood of infection. Moreover, the scale of our analysis is restricted to the city level (because our datasets are collected at this level) and so our findings do not necessarily generalize to the perhaps more important case of country-level and international travel policies. If the $r\cdot f$ invariance, or some form of inverse relation between $r$ and $f$, holds for international mobility patterns, then, as at the city scale, lax distance limits could be compensated by strict frequency restrictions. This is a bold hypothesis, which should be tested in future work. Metapopulation disease models \cite{watts2005multiscale,colizza2007invasion,arino2006disease}, with their convenient trade off between realism and parsimony, seem like a good theoretical starting point for this effort.

Despite these limitations, we believe our results reveal novel insights and bring up an important component of human mobility that could prove useful for future policies on COVID-19 and other epidemic diseases. In the current COVID-19 pandemic and in future pandemics, evidence-based policies at the city scale (not to mention the national or international scale) are needed to mitigate speed and intensity of disease spread. Our results have the potential to help this effort by opening up investigation into the relationship between exploration velocity and epidemic spread. They indicate that the exploration velocity of a city's inhabitants $r\cdot f$ must be bounded -- to bound distance $r$ but not visitation frequency $f$ leaves potential for disease containment unmet. Furthermore, and more optimistically, if a bound on exploration velocity is possible and effective, it would mean that strict distance restrictions at the city or town level -- as adopted by, e.g.,  Ireland and Italy at the beginning of the infection -- are perhaps unnecessary. Given a desired bound on epidemic speed $\tau_{bound}$, a large $r$ could be offset by a small $f$; allowing citizens to travel infrequently to distant services (doctors, hospitals etc) may be safe. This inverse relation between travel distance and frequency ($r \propto \tau_{bound} / f$) could also vitally inform remote working policies supporting the hypothesis that working from home multiple days per week --  and thereby limiting the visitation frequency to workplaces --  helps prevent the spread of disease.

\section*{Methods}
\noindent\textbf{New York City data.} Individuals' movements in New York City are inferred from GPS traces collected from mobile phones by the company X-Mode over a span of one month (February 2020).  The raw data contains about 479,163 anonymized users; our analysis uses 10,000 users randomly selected from those that appear in the dataset every day in the month of February.

\noindent\noindent\textbf{Dakar data.} The Dakar dataset is based on anonymized Call Detailed Records (CDR) provided by the Data for Development (D4D) Challenge. The detailed information of this dataset is provided in \cite{de2014d4d}. Here, we use the SET2, which includes individual trajectories for 300,000 sampled users in Senegal, and after the preprocess, we have 173,000 users and 173 cells in Dakar region during two weeks of January, 2013. We subselect for users who appear at least 200 times in the dataset to ensure that we have adequate information about their trajectories over the two weeks. 

\noindent\textbf{Data preprocessing.} The X-Mode data from NYC is generated on an very fine spatial and temporal scale, with exact latitude and longitude coordinate updates as frequently as every second. The CDR data from Dakar, on the other hand, are generated only for voice calls, text messages or data exchanges and therefore have limited resolution in time. The geographic location of the cell towers and their density determines the accuracy of location measurements through triangularization techniques. Therefore, the trajectories extracted from CDRs constitute a discrete approximation of the moving population $M(x; y; t)$. There are several steps in preprocessing of the data before it can be suitable for use in our analysis, which vary between the X-Mode data and the CDR data.

The main steps for the NYC data are:
i) We use density-based spatial clustering of applications with noise (DBSCAN) to group tightly-clustered latitude/longitude pairs in each individual's trajectory into locations \cite{ester1996density}. If a cluster of at least five latitude/longitude points exists such that no point is more than .0004 degrees (about 56 meters) from two other points in the cluster, those points are grouped together as a single location. ii) Each agent is assigned the DBSCAN cluster it visits most as its home location. iii) We drop all locations in the trajectory that have been visited for less than a minimum time $\tau_{min} = 15{\text min}$. iv) In order to restrict travel distance $r$, we calculate distance between locations by the haversine formula, which derives the great-circle distance between two points on a sphere. All locations that are more than $r~km$ from an agents home location are removed from their trajectory. v) In order to restrict travel frequency $f$, for each DBSCAN cluster that an agent visits more than $f$ distinct times (where distinct visits are determined by an agent leaving a location and then coming back to it), we randomly select $f$ visits to include in their trajectory and drop the rest.

The main steps for the Dakar data are: 
i) We view each cell tower as a different location in the city. ii) For each person, we determine the home location as the cell tower location which has been visited for the most cumulative time. By summing over all days in a given time window, one can find the home cell with high level of confidence for the majority of subjects. iii) We drop all locations in the trajectory that have been visited for less than a minimum time $\tau_{min} = 10{\text min}$.  iv) In order to restrict travel distance $r$, we calculate distance between cell towers by the haversine formula, which derives the great-circle distance between two points on a sphere. All cell towers that are more than $r~km$ from an agent's home location are removed from their trajectory. v) In order to restrict travel frequency $f$, for each location that an agent visits more than $f$ distinct times (where distinct visits are determined by an agent leaving a location and then coming back to it), we randomly select $f$ visits to include in their trajectory and drop the rest (excepting visits to an agent's home location, which are not restricted).

The duration of stay, frequency, and distance criteria on defining cell visits yields a list of cells visited by that subject over the study period for a given frequency restriction $f$ and distance restriction $r$.

\noindent\textbf{Simulation details.} We run an agent-based SEIR, SIR, and SI models with $N = 10,000$ agents, 5\% of which are initialized to be infected (1\% in the SI and SIR Dakar models). Each agent is assigned the trajectory of a real person from our dataset, with location updated every 900 seconds (15 minutes) for the NYC simulations or every 600 seconds (10 minutes) for the Dakar simulations. At each time step, each user's location is updated according to their assigned trajectory and infection status is updated according to the following parameters, drawn from Chen 2020\cite{chen2020mathematical}'s estimates of R0 = 3.58, incubation period = 5.2 days, and infection period = 5.8 days:
\begin{itemize}
\item[] $\beta = $ daily transmission parameter $ = \frac{3.58}{5.8} = .617$
\item[] $\sigma = $ daily rate at which an exposed person becomes infective $ = 1/5.2$
\item[] $\gamma = $ daily recovery parameter $ = 1/5.8$
\end{itemize}
Let $s$ be the number of time steps in a day. We then transform the above daily parameters into timestep parameters as follows:
\begin{itemize}
\item[] $\beta^* = $ time step transmission probability $ = \beta/s$
\item[] $\gamma^* = $ time step recovery probability $ = 1 - \sqrt[s]{1-\gamma} $
\item[] $\sigma^* = $ time step probability that an exposed person becomes infective $ = 1 - \sqrt[s]{1-\sigma} $
\end{itemize}
In addition, we reproduce our results with disease parameters estimated for the more highly-contagious Delta variant of COVID-19 ($\beta = 1.42$, all other parameters remain the same, \cite{Kang2021}) and the  2009 H1N1 influenza strain ($\beta = .913, \gamma = 1.6, \sigma = 1$, \cite{Biggerstaff2014,Cori2012}).
Finally, let $I_{\text{local}}$ and $N_{\text{local}}$ be the number of infected agents and total agents within a 190 meter radius of the agent's current location for the NYC data or within the same cell tower location for the Dakar data. Then, transition probabilities are:
\begin{itemize}
\item[] $\mathbb{P}[S\to E] = \beta^* * \frac{I_{\text{local}}}{N_{\text{local}}}$
\item[] $\mathbb{P}[E\to I] = \sigma^*$
\item[] $\mathbb{P}[I\to R] = \gamma^*$
\end{itemize}

 In the SIR and SI models, if an agent becomes infected on a given day, they will become contagious at the start of the next day.

  \noindent\textbf{Quantifying dispersion.} We use the M function developed in \cite{Marcon2010} to quantify spatial dispersion of disease in our New York City simulations for a given $r$, $f$. The M function is calculated as follows: for each infected agent $I$ and some radius $k$, we calculate the ratio between the proportion of agents within $k$ of $I$ which are infected to the proportion of agents in the total population which are infected. Summing this value over all infected agents $I$ and dividing by $N-1$, where $N$ is the number of infected agents, gives $M(k)$, the M function evaluated at $k$. While M is generally analyzed as a function over all reasonable $k$, we evaluate the M function at a specific $k$ in order to compare spatial dispersion across $r\cdot f$ values at that $k$, and then show that the relationship is robust to choice of $k$.
 
 Confidence intervals for $M(k)$ are obtained by Monte Carlo simulation---for a given epidemic size $\psi$, we randomly assign $\psi$ infections across the population 1,000 times and calculate $M_\psi$ each time. By taking the $.025$ and $.975$ quantiles of these simulated $M$, we form an upper and lower bound on $M_\psi$. It is notable that our empirical $M$ never reaches this confidence band, implying that spatial dispersion is significantly non-homogenous for every $k$ and $\psi$.
 
\begin{figure*}[!h]
    \centering 
\begin{subfigure}{0.45 \textwidth}
  \includegraphics[width=\linewidth]{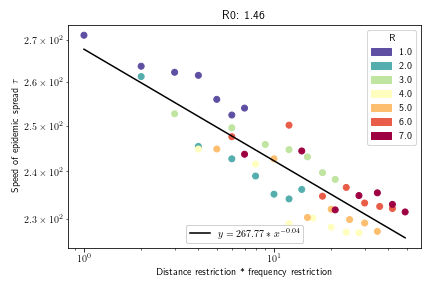}
  \label{h1n1}
\end{subfigure}\hfil 
\medskip
\begin{subfigure}{0.45\textwidth}
  \includegraphics[width=\linewidth]{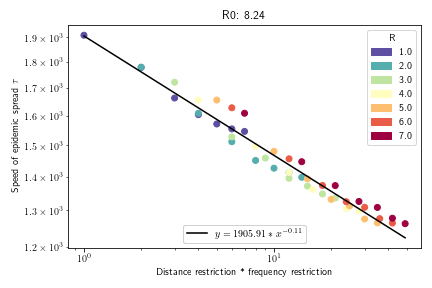}
  \label{delta}
\end{subfigure}
\caption{\textbf{Epidemics sizes for NYC covid outbreak with lower $R_0 = 1.46$ (similar to the H1N1 epidemic) and higher $R_0 = 8.2$ (similar to the upper estimates of COVID-19 Delta variant contagion), suggesting the scaling collapse is robust.} $R^2$ of best-fit lines are .925 and .9625, respectively.}
\label{robust}
\end{figure*}

 \noindent\textbf{Robustness to model parameters.} Here we demonstrate that the universal $\tau(r\cdot f)$ curve is robust to changes in the parameters $R_0$ by running our simulations with estimated transmission parameters for the 2009 H1N1 influenza strain ($\beta = .913, \gamma = 1.6, \sigma = 1$ \cite{Biggerstaff2014,Cori2012}) and the Delta variant of COVID-19 ($\beta = 1.42$, \cite{Kang2021}). Figure~\ref{robust} shows that, with these parameters, the $\tau(r\cdot f)$ relationship still holds. 

\noindent\textbf{Preferential return model.} The preferential return model proposed in \cite{Schlapfer2021} is based off of that proposed in Song et al. \cite{song2010modelling}. With some probability $P_{\text{new}}$, agents return to a location they have already visited; with probability $1-P_{\text{new}}$ they visit a new location with distance drawn from the empirical distance distribution and direction drawn uniformly at random. Waiting times between trips are also drawn from the empirical distribution. The probability $p$ is a function of the number of locations already visited:
$$P_{\text{new}} = \rho S^{-\gamma}$$
 where $S$ is the number of locations visited. The parameters $\rho$ and $\gamma$ are fit to the real data using least-squares regression (using the NYC dataset, we find $\rho = .500$ and $\gamma = .267$.)

We add the following modification: while in the original PEPR model agents travel at each timestep, we set some probability $P_{\text{travel}}$ with which agents travel at any given timestep. Conversely, with probability $1-P_{\text{travel}}$, agents stay home. The simulation results reported here are from running our modified PEPR on a unit square for 500 timesteps.





  \section*{Acknowledgments}
  The authors would like to thank Allianz,  Amsterdam Institute for Advanced Metropolitan Solutions, Brose, Cisco, Ericsson, Fraunhofer Institute, Liberty Mutual Institute, Kuwait-MIT Center for Natural Resources and the Environment, Shenzhen, Singapore- MIT Alliance for Research and Technology (SMART), UBER, Vitoria State Government, Volkswagen Group America, and all the members of the MIT Senseable City Lab Consortium for supporting this research. Research of S.H.S. was supported by NSF Grants DMS-1513179 and CCF-1522054. The authors would also like to thank X-Mode Social, Inc., for provision of the New York City data. We would like to acknowledge ORANGE / SONATEL for providing the data.
\section*{Author contributions}
    C.R. conceived the work. K.P.O. and C.H. designed the simulations. L.Y. acquired the data. C.H. executed the simulations. K.P.O., P.S., and C.H. contributed to writing and revising the manuscript. P.S. and C.R. supervised the research.
\section*{Data availability}
The data and code used in this study are available from the authors upon reasonable request and with permission of X-Mode Social, Inc.
\bibliographystyle{naturemag}
\bibliography{ref.bib}

\begin{thebibliography}{10}
\expandafter\ifx\csname url\endcsname\relax
  \def\url#1{\texttt{#1}}\fi
\expandafter\ifx\csname urlprefix\endcsname\relax\def\urlprefix{URL }\fi
\providecommand{\bibinfo}[2]{#2}
\providecommand{\eprint}[2][]{\url{#2}}

\bibitem{wilson1995travel}
\bibinfo{author}{Wilson, M.~E.}
\newblock \bibinfo{title}{Travel and the emergence of infectious diseases.}
\newblock \emph{\bibinfo{journal}{Emerging infectious diseases}}
  \textbf{\bibinfo{volume}{1}}, \bibinfo{pages}{39} (\bibinfo{year}{1995}).

\bibitem{sattenspiel1995}
\bibinfo{author}{Sattenspiel, L.} \& \bibinfo{author}{Dietz, K.}
\newblock \bibinfo{title}{A structured epidemic model incorporating geographic
  mobility among regions}.
\newblock \emph{\bibinfo{journal}{Mathematical Biosciences}}
  \textbf{\bibinfo{volume}{128}}, \bibinfo{pages}{71--91}
  (\bibinfo{year}{1995}).
\newblock
  \urlprefix\url{https://www.sciencedirect.com/science/article/pii/002555649400068B}.

\bibitem{dengue2015}
\bibinfo{author}{Wesolowski, A.} \emph{et~al.}
\newblock \bibinfo{title}{Impact of human mobility on the emergence of dengue
  epidemics in pakistan}.
\newblock \emph{\bibinfo{journal}{Proceedings of the National Academy of
  Sciences}} \textbf{\bibinfo{volume}{112}}, \bibinfo{pages}{11887--11892}
  (\bibinfo{year}{2015}).
\newblock \urlprefix\url{https://www.pnas.org/doi/abs/10.1073/pnas.1504964112}.
\newblock \eprint{https://www.pnas.org/doi/pdf/10.1073/pnas.1504964112}.

\bibitem{flu2017}
\bibinfo{author}{Charu, V.} \emph{et~al.}
\newblock \bibinfo{title}{Human mobility and the spatial transmission of
  influenza in the united states}.
\newblock \emph{\bibinfo{journal}{PLOS Computational Biology}}
  \textbf{\bibinfo{volume}{13}}, \bibinfo{pages}{1--23} (\bibinfo{year}{2017}).
\newblock \urlprefix\url{https://doi.org/10.1371/journal.pcbi.1005382}.

\bibitem{barmak2016modelling}
\bibinfo{author}{Barmak, D.~H.}, \bibinfo{author}{Dorso, C.~O.} \&
  \bibinfo{author}{Otero, M.}
\newblock \bibinfo{title}{Modelling dengue epidemic spreading with human
  mobility}.
\newblock \emph{\bibinfo{journal}{Physica A: Statistical Mechanics and its
  Applications}} \textbf{\bibinfo{volume}{447}}, \bibinfo{pages}{129--140}
  (\bibinfo{year}{2016}).

\bibitem{Schlapfer2021}
\bibinfo{author}{Schl{\"{a}}pfer, M.} \emph{et~al.}
\newblock \bibinfo{title}{{The universal visitation law of human mobility}}.
\newblock \emph{\bibinfo{journal}{Nature}} \textbf{\bibinfo{volume}{593}},
  \bibinfo{pages}{522--527} (\bibinfo{year}{2021}).
\newblock \urlprefix\url{https://doi.org/10.1038/s41586-021-03480-9}.

\bibitem{chen2020mathematical}
\bibinfo{author}{Chen, T.-M.} \emph{et~al.}
\newblock \bibinfo{title}{A mathematical model for simulating the phase-based
  transmissibility of a novel coronavirus}.
\newblock \emph{\bibinfo{journal}{Infectious diseases of poverty}}
  \textbf{\bibinfo{volume}{9}}, \bibinfo{pages}{1--8} (\bibinfo{year}{2020}).

\bibitem{Marcon2010}
\bibinfo{author}{Marcon, E.} \& \bibinfo{author}{Puech, F.}
\newblock \bibinfo{title}{{Measures of the geographic concentration of
  industries: improving distance-based methods}}.
\newblock \emph{\bibinfo{journal}{Journal of Economic Geography}}
  \textbf{\bibinfo{volume}{10}}, \bibinfo{pages}{745--762}
  (\bibinfo{year}{2010}).
\newblock \urlprefix\url{https://doi.org/10.1093/jeg/lbp056}.

\bibitem{cirillo2020tail}
\bibinfo{author}{Cirillo, P.} \& \bibinfo{author}{Taleb, N.~N.}
\newblock \bibinfo{title}{Tail risk of contagious diseases}.
\newblock \emph{\bibinfo{journal}{Nature Physics}} \bibinfo{pages}{1--8}
  (\bibinfo{year}{2020}).

\bibitem{donnat2020modeling}
\bibinfo{author}{Donnat, C.} \& \bibinfo{author}{Holmes, S.}
\newblock \bibinfo{title}{Modeling the heterogeneity in covid-19's reproductive
  number and its impact on predictive scenarios}.
\newblock \emph{\bibinfo{journal}{Journal of Applied Statistics}}
  \textbf{\bibinfo{volume}{0}}, \bibinfo{pages}{1--29} (\bibinfo{year}{2021}).
\newblock \urlprefix\url{https://doi.org/10.1080/02664763.2021.1941806}.
\newblock \eprint{https://doi.org/10.1080/02664763.2021.1941806}.

\bibitem{Buchanan2020limits}
\bibinfo{author}{Buchanan, B.}
\newblock \bibinfo{title}{The limits of a model}.
\newblock \emph{\bibinfo{journal}{Nature Physics}} \bibinfo{pages}{1--8}
  (\bibinfo{year}{2020}).

\bibitem{ma2020epidemiological}
\bibinfo{author}{Ma, S.} \emph{et~al.}
\newblock \bibinfo{title}{Epidemiological parameters of coronavirus disease
  2019: a pooled analysis of publicly reported individual data of 1155 cases
  from seven countries}.
\newblock \emph{\bibinfo{journal}{Medrxiv}}  (\bibinfo{year}{2020}).

\bibitem{watts2005multiscale}
\bibinfo{author}{Watts, D.~J.}, \bibinfo{author}{Muhamad, R.},
  \bibinfo{author}{Medina, D.~C.} \& \bibinfo{author}{Dodds, P.~S.}
\newblock \bibinfo{title}{Multiscale, resurgent epidemics in a hierarchical
  metapopulation model}.
\newblock \emph{\bibinfo{journal}{Proceedings of the National Academy of
  Sciences}} \textbf{\bibinfo{volume}{102}}, \bibinfo{pages}{11157--11162}
  (\bibinfo{year}{2005}).

\bibitem{colizza2007invasion}
\bibinfo{author}{Colizza, V.} \& \bibinfo{author}{Vespignani, A.}
\newblock \bibinfo{title}{Invasion threshold in heterogeneous metapopulation
  networks}.
\newblock \emph{\bibinfo{journal}{Physical review letters}}
  \textbf{\bibinfo{volume}{99}}, \bibinfo{pages}{148701}
  (\bibinfo{year}{2007}).

\bibitem{arino2006disease}
\bibinfo{author}{Arino, J.} \& \bibinfo{author}{Van~den Driessche, P.}
\newblock \bibinfo{title}{Disease spread in metapopulations}.
\newblock \emph{\bibinfo{journal}{Fields Institute Communications}}
  \textbf{\bibinfo{volume}{48}}, \bibinfo{pages}{1--12} (\bibinfo{year}{2006}).

\bibitem{de2014d4d}
\bibinfo{author}{de~Montjoye, Y.-A.}, \bibinfo{author}{Smoreda, Z.},
  \bibinfo{author}{Trinquart, R.}, \bibinfo{author}{Ziemlicki, C.} \&
  \bibinfo{author}{Blondel, V.~D.}
\newblock \bibinfo{title}{D4d-senegal: The second mobile phone data for
  development challenge} (\bibinfo{year}{2014}).
\newblock \eprint{1407.4885}.

\bibitem{ester1996density}
\bibinfo{author}{Ester, M.}, \bibinfo{author}{Kriegel, H.-P.},
  \bibinfo{author}{Sander, J.} \& \bibinfo{author}{Xu, X.}
\newblock \bibinfo{title}{A density-based algorithm for discovering clusters in
  large spatial databases with noise}.
\newblock In \emph{\bibinfo{booktitle}{Proceedings of the Second International
  Conference on Knowledge Discovery and Data Mining}},
  \bibinfo{pages}{226--231} (\bibinfo{year}{1996}).

\bibitem{Kang2021}
\bibinfo{author}{Kang, M.} \emph{et~al.}
\newblock \bibinfo{title}{Transmission dynamics and epidemiological
  characteristics of delta variant infections in china}.
\newblock \emph{\bibinfo{journal}{medRxiv}}  (\bibinfo{year}{2021}).
\newblock
  \urlprefix\url{https://www.medrxiv.org/content/early/2021/08/13/2021.08.12.21261991}.
\newblock
  \eprint{https://www.medrxiv.org/content/early/2021/08/13/2021.08.12.21261991.full.pdf}.

\bibitem{Biggerstaff2014}
\bibinfo{author}{Biggerstaff, M.}, \bibinfo{author}{Cauchemez, S.},
  \bibinfo{author}{Reed, C.}, \bibinfo{author}{Gambhir, M.} \&
  \bibinfo{author}{Finelli, L.}
\newblock \bibinfo{title}{{Estimates of the reproduction number for seasonal,
  pandemic, and zoonotic influenza: a systematic review of the literature}}.
\newblock \emph{\bibinfo{journal}{BMC Infectious Diseases}}
  \textbf{\bibinfo{volume}{14}}, \bibinfo{pages}{480} (\bibinfo{year}{2014}).
\newblock \urlprefix\url{https://doi.org/10.1186/1471-2334-14-480}.

\bibitem{Cori2012}
\bibinfo{author}{Cori, A.} \emph{et~al.}
\newblock \bibinfo{title}{{Estimating influenza latency and infectious period
  durations using viral excretion data}}.
\newblock \emph{\bibinfo{journal}{Epidemics}} \textbf{\bibinfo{volume}{4}},
  \bibinfo{pages}{132--138} (\bibinfo{year}{2012}).
\newblock
  \urlprefix\url{http://www.sciencedirect.com/science/article/pii/S175543651200031X}.

\bibitem{song2010modelling}
\bibinfo{author}{Song, C.}, \bibinfo{author}{Koren, T.}, \bibinfo{author}{Wang,
  P.} \& \bibinfo{author}{Barab{\'a}si, A.-L.}
\newblock \bibinfo{title}{Modelling the scaling properties of human mobility}.
\newblock \emph{\bibinfo{journal}{Nature Physics}}
  \textbf{\bibinfo{volume}{6}}, \bibinfo{pages}{818} (\bibinfo{year}{2010}).

\end{thebibliography}
\pagebreak
\begin{figure*}[htb]
    \centering 
\begin{subfigure}{0.45 \textwidth}
  \includegraphics[width=\linewidth]{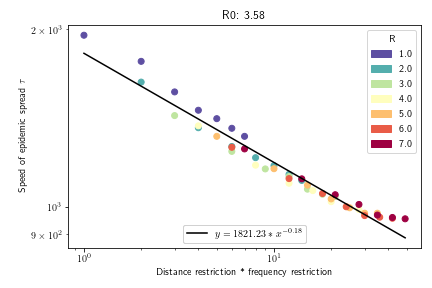}
  \caption{NYC}
  \label{rfsinyc}
\end{subfigure}\hfil 
\begin{subfigure}{0.45 \textwidth}
  \includegraphics[width=\linewidth]{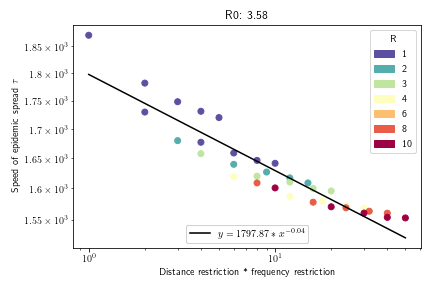}
  \caption{Dakar}
  \label{rfsidak}
\end{subfigure}\hfil 

\medskip
\begin{subfigure}{0.45\textwidth}
  \includegraphics[width=\linewidth]{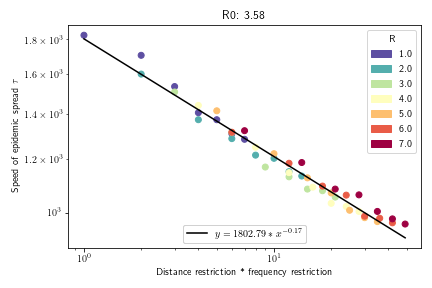}
  \caption{NYC}
  \label{rtimesfsirnyc}
\end{subfigure}\hfil 
\begin{subfigure}{0.45\textwidth}
  \includegraphics[width=\linewidth]{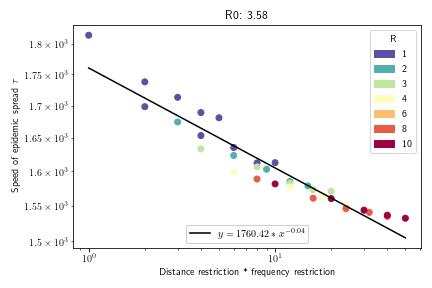}
  \caption{Dakar}
  \label{rtimesfsirdak}
\end{subfigure}\hfil 
\caption{\textbf{Scaling collapse in SI and SIR model}. Top row: scaling collapse for SI model. $R^2$ values for best-fit lines are, from left to right, .958 and .896.  Best-fit line parameters are $a = -0.18, b = 1821.23$ (NYC) and $a = -0.04, b = 1797.87$ (Dakar). Bottom row, scaling collapse for SIR model. $R^2$ values for best-fit lines are, from left to right, .974 and .937. Best-fit line parameters are $a = -0.17, b = 1802.79$ (NYC) and $a = -0.04, b = 1760.42$ (Dakar).}
\label{SISIR}
\end{figure*}

\begin{figure*}
\begin{subfigure}{0.9\textwidth}
  \includegraphics[width=\columnwidth]{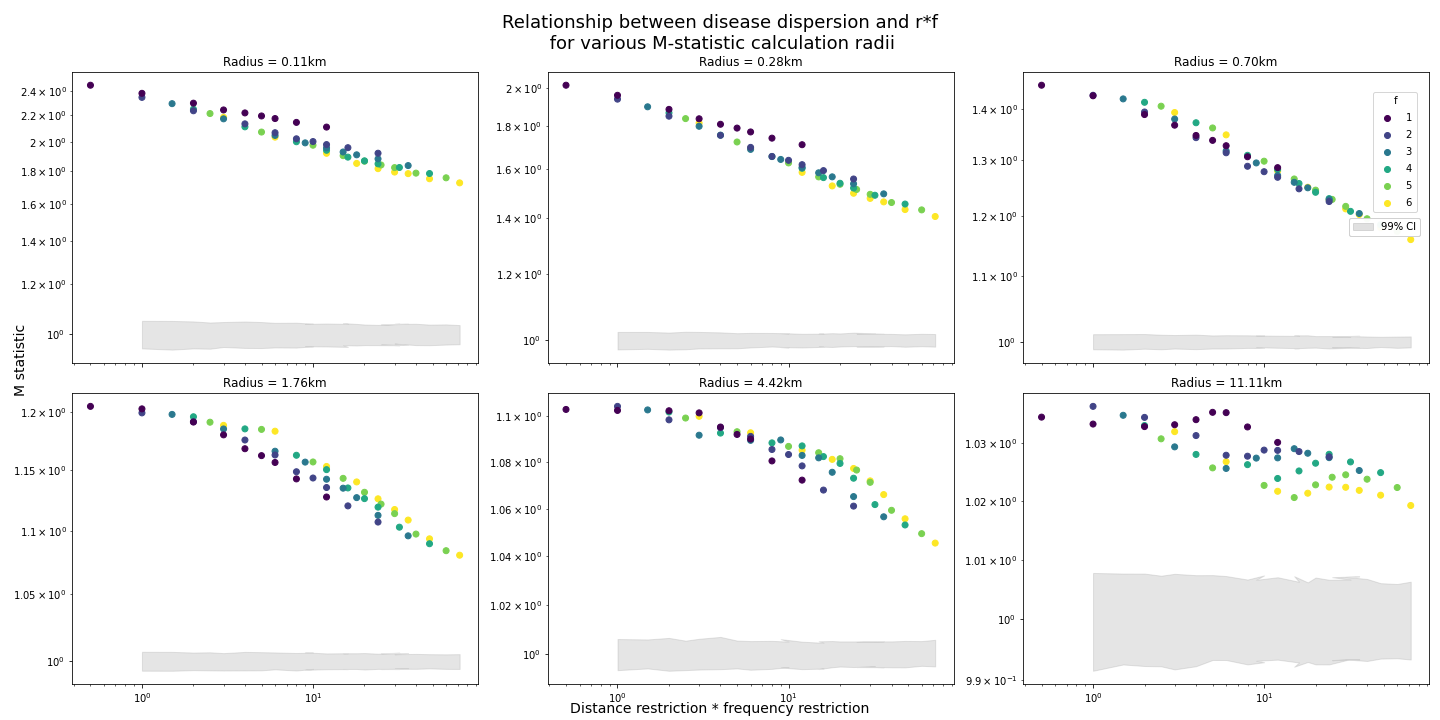}
\end{subfigure}\hfil 
\caption{\textbf{Collapse of spatial dispersion of infections for varius $k$.} Spatial dispersion $M(k)$ shows a scaling relationship with $r\cdot f$ regardless of $k$. 99\% confidence bands are shown in gray, indicating that the spatial clustering in infections remains significant across values of $r\cdot f$.}
\label{fig:spatialcollapse}
\end{figure*} 

\end{document}